\documentclass[a4paper,12pt]{article}
\usepackage{graphicx}
\begin{document}

\title{Has Dark Energy really been discovered in the Lab?}

\author{Philippe Jetzer and Norbert Straumann\\
        Institute for Theoretical Physics University of Zurich,\\
        CH--8057 Zurich, Switzerland}
\maketitle

\begin{abstract}
We show that Dark Energy contributions can \textit{not} be determined from noise
measurements of Josephson junctions, as was recently suggested in a paper by C. Beck and
M.C. Mackey.
\end{abstract}

\section*{Zero-point energies and non-gravitational\\ effects}

In a recent paper \cite{1} of C. Beck and M.C. Mackey the possibility was suggested that
part of the zero-point energy of the radiation field that is gravitationally active can be
determined  from noise measurements of Josephson junctions. We are convinced that there is
no basis for this. The reason being that the absolute value of the zero-point energy of a
quantum mechanical system has no physical meaning when gravitational coupling is ignored.
All that is measurable are changes of the zero-point energy under variations of system
parameters or of external couplings, like an applied voltage. This was nicely illustrated
some time ago by C. Enz and A. Thellung in a paper \cite{2} contained in the memorial
volume of Helv. Phys. Acta for W. Pauli. Beside the Casimir effect, these authors
considered the sublimation pressure of isotopes. Contrary to earlier beliefs of O. Stern
and W. Pauli \cite{3}, the latter is independent of whether the zero-point energies of the
lattice vibrations of the solid are included or not, as long as the total Hamiltonian is
symmetric in all the atoms (in the solid and vapor phases).

The simplest example which illustrates our point is the following caricature for the van
der Waals force. Consider two identical harmonic oscillators of frequency $\omega_0$
separated by a distance $R$, which are harmonically coupled with the usual dipole-dipole
interaction strength $(e^2/R^3)q_1q_2$. With a simple canonical transformation we can
decouple the two harmonic oscillators, and the ground state energy of the system is just
the sum of the zero-point energies of the decoupled oscillators. The corresponding van der
Waals force is, however, independent of the zero-point energies of the original
oscillators. The latter is not measurable without considering gravitational coupling.

The same holds for the more complicated example of C. Beck and M. Mackey. Their split of
the spectral density for the junction noise current into a ``vacuum and ordinary
Bose-Einstein parts'' is formal. Indeed, the spectral density originally comes from a
simple rational expression of Boltzmann factors, which are not related to zero-point
energies. This kind of laboratory experiment can not shed light on the nature of the dark
energy.

In order to demonstrate the last claim we consider again our toy model, this time at finite
temperatures. The van der Waals force $K$ is given by the derivative of the free energy
$F$,
\begin{equation}
K=-\frac{\partial F}{\partial R},~~~F=-k_BT\ln Z,
\end{equation}
with
\begin{equation}
Z=\prod_{i=1,2}e^{-\hbar\omega_i/2k_BT}\frac{1}{1-e^{-\hbar\omega_i/k_BT}},
\end{equation}
where $\omega_i(R)$ are the two frequencies of the decoupled oscillators
($\omega_i^2=\omega_0^2\pm e^2/(mR^3)$). Hence the van der Waals force is
\begin{equation}
K=-\sum_{i=1,2}\left[
\frac{\hbar\omega_i}{2}+\frac{\hbar\omega_i}{e^{\hbar\omega_i/k_BT-1}}\right]
\frac{\partial\omega_i/\partial R}{\omega_i},
\end{equation}
or
\begin{equation}
K \simeq -2 \left[
\frac{\hbar\omega_0}{2}+\frac{\hbar\omega_0}{e^{\hbar\omega_0/k_BT-1}}\right]
 \sum_{i=1,2} \frac{1}{2}\frac{\partial(\omega_i/\omega_0)}{\partial R}.
\end{equation}
The spectral function for the response to the change of the distance $R$ is thus
\textit{exactly the same} as in the paper of C. Beck and M. Mackey, but obviously the
``vacuum part'' in the square bracket has nothing to do with the zero-point energies of the
two uncoupled oscillators, because the latter does not contribute to the van der Waals
force. The formal identity of the two is really misleading. We could have used normal
ordering for the Hamiltonian of the unperturbed harmonic oscillators without changing the
result (3). We hope that this simple consideration clarifies the issue, which received
widespread attention in  semi-popular  and other journals \cite{4}.


\begin{thebibliography}{99}
\bibitem{1}
C. Beck and M.C. Mackey, astro-ph/0406504, to appear in Phys. Lett. B.
\bibitem{2}
C.P. Enz and A. Thellung, \textit{Helv. Phys. Acta}, \textbf{33}, 839 (1960); see also N.
Straumann, {\it On the Cosmological Constant Problems and the Astronomical Evidence for a
Homogeneous Energy Density with Negative Pressure}, in {\it Poincar\`{e} Seminar 2002,
Vacuum Energy -- Renormalization},  B. Duplantier, and V. Rivasseau, eds.;
Birkh\"{a}user-Verlag 2003, p.7-51; astro-ph/0203330.
\bibitem{3}
W. Pauli, {\it Pauli Lectures on Physics}; Ed. C.P. Enz. MIT Press (1973); Vol. 4,
especially Sect. 20.
\bibitem{4}
P. Ball, \textit{Nature}, news item, vol. \textbf{430}, 126 (2004).

\end{thebibliography}
\end{document}